\documentclass[amsmath,amssymb,twocolumn,prb]{revtex4}

\ifx\pdftexversion\undefined
  \usepackage[dvips]{graphics}
\else
  \usepackage[pdftex]{graphics}
\fi

\usepackage{graphicx}

\begin{document}

\title{Spin-photovoltaic effect in quantum wires due to inter-subband transitions}

\author{Arkady Fedorov} \email{fedorov@clarkson.edu}
\affiliation{Center for Quantum Device Technology, Department of
Physics and Department of Electrical and Computer Engineering,
Clarkson University, Potsdam, NY 13699-5721, USA}

\author{Yuriy V. Pershin} \email{pershin@pa.msu.edu} \author{Carlo
Piermarocchi} \email{piermarocchi@pa.msu.edu} \affiliation{Department
of Physics and Astronomy, Michigan State University, East Lansing,
Michigan 48824-2320, USA}

\begin{abstract}
We consider the current induced in a quantum wire by external
electromagnetic radiation. The photocurrent is caused by the
interplay of spin-orbit interaction (Rashba and Dresselhaus terms)
and external in-plane magnetic field. We calculate this current
using a Wigner functions approach taking into account
radiation-induced transitions between transverse subbands. The
magnitude and the direction of the current depend on the
Dresselhaus and Rashba constants, strength of magnetic field,
radiation frequency and intensity. The current can be controlled
by changing some of these parameters.
\end{abstract}

\pacs{73.21.Hb, 72.25.Dc, 71.70.Ej, 72.40.+w, 72.30.+q} \maketitle

\section{Introduction}
The energy spectrum of free electrons in a perfect quantum wire
without spin-orbit interaction consists of spin-degenerate subbands
due to the transverse confinement in two directions. In each subband,
the energy depends quadratically on the one-dimensional momentum. In
the presence of an external radiation intersubband excitation
probabilities are equal for states with opposite momentum. Therefore,
there is no change in current associated with external radiation.

During the last 15 years there has been a great interest in
theoretical and experimental investigations of photovoltaic
effects and photoconductance in quantum wires (see Refs.
\onlinecite{lit2}-\onlinecite{lit} and references therein).
Mechanisms for pure spin current generation in 2D and 1D systems
with spin-orbit interaction have also been discussed
~\cite{sipe,ivchenko}. It is known that a photovoltaic effect is
possible in quantum wires without inversion symmetry. For example,
a photovoltaic effect in quantum wires with spatially dependent
lateral confinement was predicted in Ref.~\onlinecite{lenya}. In
the present paper we consider a different {\it spin}-based
mechanism for photovoltaic effect, which is very interesting in
the context of the fast growing field of
spintronics.~\cite{book,review,ourrev} The most important
component of our scheme is the spin-orbit interaction. However,
the spin-orbit interaction alone is not sufficient to generate a
charge photocurrent if the quantum wire is spatially homogeneous.
Therefore, we consider a wire in an in-plane magnetic field which
breaks the inversion symmetry.

We can identify the following groups of inter-subband transitions that
lead to a photovoltaic effect in quantum wires:
\begin{enumerate}
\item Transitions between spin-splitted subbands with the same
confinement quantum numbers.
\item Transitions between subbands with different confinement quantum
numbers.
\end{enumerate}
The main difference between these two groups is that the first is
generated by the magnetic field component of electromagnetic
radiation, while transitions from the second group are due to the
electric field component. In a recent paper~\cite{perpier} a
spin-photovoltaic effect in quantum wires due to transitions of the
first type was considered.  It was found that a special role in the
photovoltaic effect is played by transitions in which the direction of
electron velocity changes. The importance of this velocity inversion
was outlined earlier, see e.g. Ref. \onlinecite{lit4}, in studies of
photoconductance. Here we consider intersubband transitions of the
second type.  An important feature in semiconductor-based quantum
wires is that the spin-orbit interaction constants are different for
subbands with different confinement quantum numbers. This peculiarity
is essential in our scheme for the generation of photocurrent.

The goal of this paper is to calculate the current in a quantum wire
at zero bias voltage due to external radiation. The current as a
function of radiation frequency is found numerically from coupled
equations involving Wigner functions. The Wigner function formalism
has many advantages for investigating transport
problems~\cite{Kluksdahl, Frensley}. Among them we mention the
phase-space nature of Wigner functions which are similar to the
classical Boltzmann distribution functions. This feature makes
possible to separate the incoming and outgoing components of the
electron distribution at the boundaries which, in turn, facilitates
the modeling of an ideal contact. The commonly used assumptions are:
the distribution of electrons emitted in the quantum wire can be
described by the equilibrium distribution function of the leads
reservoirs, and all electron are collected by the leads reservoirs
without reflection. In this work we extend the description of the
transport dynamics to include inter-subband transitions due to
electro-magnetic wave excitation.

We show that the current is sensitive to many control parameters like,
e.g., the spin-orbit coupling and external magnetic field.  Therefore,
the current can be used to determine materials parameters.  The
calculated current strength for a realistic set of parameters is of
the order of $0.1$ nA and consequently can be measured using standard
experimental techniques.

This paper is organized as follows. The single particle energy
spectrum and wave functions are introduced in Sec. \ref{sec2}. A set
of coupled equations for Wigner functions is derived in Sec.
\ref{sec3}. The discrete model used for numerical solutions is
presented in Sec. \ref{discrete}. Numerical results are given and
discussed in Sec. \ref{sec4} and concluding remarks are in Sec.
\ref{sec5}.

\section{System} \label{sec2}
\begin{figure}[bt]
\centering
\includegraphics[angle=270,width=8.5cm]{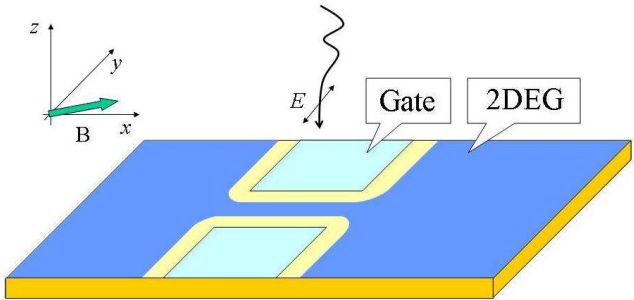}
\caption{(Color online) Quantum wire with an applied magnetic field in
($x,y$) plane, and irradiated by a electro-magnetic wave linearly
polarized in the $y$-direction.} \label{fig1}
\end{figure}

Fig. \ref{fig1} shows a possible experimental realization of the
system under investigation. The two-dimensional electron gas is split
into two parts by a potential applied to the gate electrodes. The
narrow channel between the gates then forms a quantum wire. Let us
define a coordinate system such that the direction of the electron
transport through the wire is in the $x$-direction and the lateral
confinement is in the $y$-direction. We assume that an external
magnetic field is applied in the ($x,y$) plane. Previously, several
interesting investigations of quantum wires with spin-orbit
interaction in the presence of an in-plane magnetic field were
reported.~\cite{QWSOMF1,QWSOMF2,QWSOMF3,QWSOMF5,QWSOMF6}

In the quantum wire, the Hamiltonian for the conduction electrons can
be written in the form
\begin{equation}
H=H_0+H_1,\label{ham}~,
\end{equation}
where
\begin{equation}
H_0=\frac{\mathbf{p}^2}{2m^*}-\frac{\alpha}{\hbar}p_x\sigma_y-\frac{\beta}{\hbar}p_x\sigma_x
+V(y)+U(z)+\frac{g^*\mu_B}{2}\boldsymbol{\sigma B}, \label{ham0}
\end{equation}
and
\begin{equation}
H_1= -\frac{e}{m^*}\mathbf{A}\mathbf{p}=-\frac{eE_y
p_y}{m^*\omega}\cos(\omega t ). \label{ham1}
\end{equation}
Here, $H_0$ is the time-independent part of the Hamiltonian, $H_1$
describes the interaction with the electro-magnetic field,
$\mathbf{p}$ is the momentum of the electron, $m^*$ is the effective
mass, $V(y)$ is the lateral confinement potential due to the gates,
$U(z)$ is the confinement potential in $z$-direction, $\mu_B$ and
$g^*$ are the Bohr magneton and effective g-factor, $E_y$ and $\omega$
are the amplitude and frequency of electric field of the polarized
electro-magnetic wave, and $\boldsymbol{\sigma}$ is the vector of the
Pauli matrices. The effect of the external field $\mathbf{B}$ on the
spatial motion is neglected, assuming strong confinement in the
$z$-direction. The second and third terms in (\ref{ham}) represent the
Rashba and Dresselhaus spin-orbit interaction~\cite{rashba,dressel}
for an electron moving in the $x$-direction, $\alpha$ and $\beta$ are
the corresponding coupling constants.

The spin-orbit interactions included into the Hamiltonian
(\ref{ham0}) originate from bulk inversion asymmetry (giving rise
to a Dresselhaus interaction \cite{dressel}) and structural
inversion asymmetry (giving rise to a Rashba interaction
\cite{rashba}). It is well known that the spin-orbit interaction
constants are different for electrons in different transverse
subbands \cite{sipe,pramanik}. In our model we assume that the
Rashba spin-orbit interaction constant $\alpha$ depends on the
index $m$ and the Dresselhaus spin-orbit interaction constant
$\beta$ depends both on $n$ and $m$, where $n=0,1,...$ and
$m=0,1,..$ are subband indices due to confinement in $y$ and $z$
directions respectively.~\cite{sipe,pramanik} In the model of
rigid quantum wire walls $\beta=\gamma\left( \left(\pi n /
W_z\right)^2-\left(\pi m / W_y\right)^2\right)$, where $\gamma$ is
a constant.

\begin{figure}[bt]
\centering
\includegraphics[angle=270,width=8.5cm]{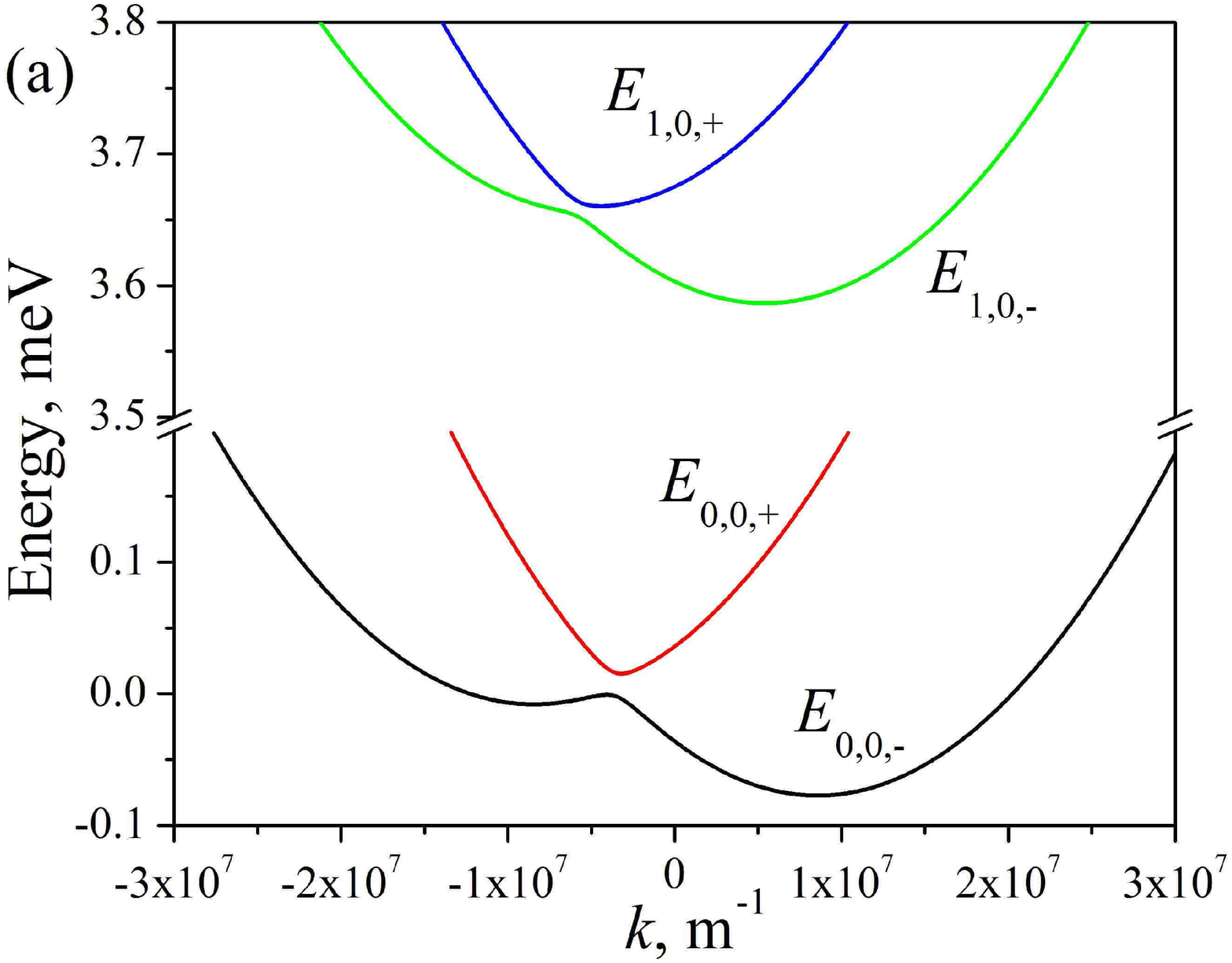}
\includegraphics[angle=270,width=8.5cm]{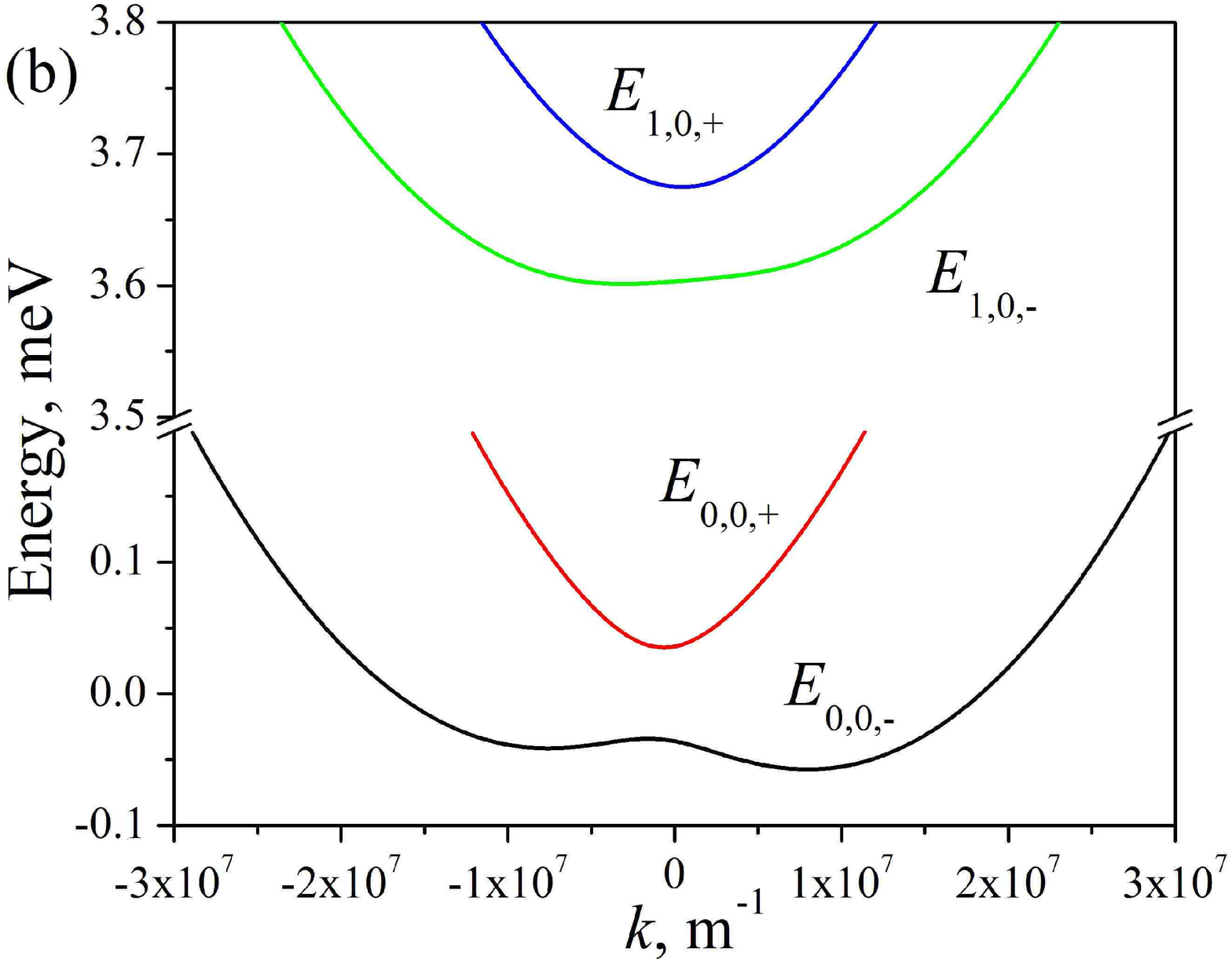}
\caption{(Color online) Dispersion relations (two lowest
spin-splitted subbands) calculated for (a) $B_x=2$T, $B_y=2$T and
(b) $B_x=2$T, $B_y=-2$T. These plots were obtained using the
parameters values: $m^*=0.067m_e$, $g^*=-0.44$,
$\alpha_0=0.5\times10^{-11}$eV m,
$\beta_{0,0}=0.84\times10^{-11}$eV m,
$\beta_{1,0}=0.36\times10^{-11}$eV m, $\epsilon_1-\epsilon_0=
3.64$meV.} \label{fig2}
\end{figure}

At $E_y = 0$, the solutions of the Schr\"odinger equation can be
written in the form

\begin{equation}
\Psi_{m,n,\pm}(k)=\frac{e^{ikx}}{\sqrt{2}}\binom{\pm
e^{i\varphi_{n,m}}}{1}\phi_m(y)\eta_n(z), \label{wf}
\end{equation}
where
\begin{eqnarray}
\varphi_{n,m}=\pi\theta(\beta_{n,m} k-\frac{g^* \mu_B}{2} B_x)+
\nonumber \\ \textnormal{arctan}\left[-\frac{-\alpha_n k+\frac{g^*
\mu_B}{2} B_y}{-\beta_{m,n} k+\frac{g^* \mu_B}{2} B_x}\right],
\label{phi}
\end{eqnarray}
$\theta(..)$ is the step function, $\phi_m(y)$ and $\eta_n(z)$ are
the wave function of the transverse modes (due to the confinement
potentials $V(y)$ and $U(z)$). The eigenvalue problem can be
solved to obtain

\begin{eqnarray}
E_{m,n,\pm}(k)=\frac{\hbar^2 k^2}{2m^*}+\epsilon_m+E_n\pm
\nonumber
\\ \sqrt{\left(-\alpha_n k+\frac{g^* \mu_B}{2}
B_y\right)^2+\left(-\beta_{n,m} k+\frac{g^* \mu_B}{2}
B_x\right)^2}. \label{spectr}
\end{eqnarray}
In this expression, $\epsilon_m$ and $E_n$ are the eigenvalues of
decoupled Scr\"odinger equations in $y$ and $z$ directions. In the
experimental setup depicted in Fig. \ref{fig1} the confinement in
$z$-direction is stronger than the confinement in $y$-direction,
thus we will consider $E_1-E_0\gg\epsilon_1-\epsilon_0$.

In what follows we assume that in the absence of radiation the
chemical potential in the wire is located between the ground
($0,0$) and first ($1,0$) transverse subbands so that only the
ground subband is occupied by electrons, and we focus our
attention on radiation-induced transitions between these two
transverse subbands. The energy spectrum in Eq. (\ref{spectr}) is
illustrated in Fig. \ref{fig2} for various directions of
$\boldsymbol{B}$. Notice that the energy dispersion is different
(not simply shifted) for the ground and the first transverse
subbands. Fig. \ref{fig2} shows that the energy spectrum is
strongly asymmetric and significantly dependent on the magnetic
field direction. We also note that the gaps between spin-splitted
subbands are due to the magnetic field.

The electron velocity is defined by the slope of $E_{m,n,\pm}(k)$
and is given by

\begin{eqnarray}
v_{n,m,\pm}(k)=\frac{1}{\hbar}\frac{\partial
E_{n,m,\pm}(k)}{\partial k}=\frac{\hbar k}{m^*}\pm \nonumber \\
\frac{\alpha_n \left(\alpha_n k-\frac{g^* \mu_B}{2}
B_y\right)+\beta_{n,m} \left(\beta_{n,m} k-\frac{g^* \mu_B}{2}
B_x\right)}{\hbar \sqrt{\left(-\alpha_n k+\frac{g^* \mu_B}{2}
B_y\right)^2+\left(-\beta_{n,m} k+\frac{g^* \mu_B}{2}
B_x\right)^2}}. \label{velocity}
\end{eqnarray}
We emphasize that the direction of velocity changes in local
extrema points of the spectrum. The radiation-induced transitions
that we will consider conserve $k$. Since the positions of local
extrema are different in the ground ($0,0$) and first ($1,0$)
transverse subbands, transitions reversing the velocity direction
are possible. The spectrum asymmetry results in asymmetric
transition rates and as a result in a finite current at zero bias
voltage.

\section{Wigner functions} \label{sec3}
\subsection{Interaction Hamiltonian}
Assuming a parabolic confinement in $y$ direction we write the
ground and the first excited wave functions of the corresponding
transverse mode explicitly as
\begin{eqnarray}\label{harmonic}
\phi_0 (y) &=& \left( \frac{2}{\pi\bar y^2 } \right)^{ - 1/4} \exp
(\frac{-y^2}{\bar y^2}),\nonumber\\
\phi_1 (y) &=&2 \left( \frac{2}{\pi\bar y^6 } \right)^{ - 1/4} y
\exp (\frac{-y^2}{\bar y^2}),
\end{eqnarray}
where $\bar y$ is the characteristic width of the quantum wire in
the $y$ direction. The energy gap between the ground and the first
excited states can be estimated as
$\epsilon_1-\epsilon_0=(2\hbar^2)/(m^* {\bar y}^2)$. Taking the
form of the solution for the Schr\"odinger equation (\ref{phi})
and (\ref{harmonic}) we obtain the matrix form of the interaction
Hamiltonian (\ref{ham1})
\begin{equation}\label{ham1t}
H_1=i\hbar g \left( {e^{i\omega t} + e^{ - i\omega t} }
\right)\left( {\sigma ^ + s^{01}  - h.c. } \right),
\end{equation}
where  $g=(\hbar eE_y) /(2m^* \bar y \omega)$ is the coupling
constant, $\sigma^{\pm}$ is the ladder operator acting on the $y$
component of the wave function:
$\sigma^{\pm}\phi_{m}(y)=\phi_{m\pm1}(y)$ and $s^{01}$ is an
operator in the space of spin degree of freedom,
\begin{equation}\label{s12}
    s^{01}=\left(\begin{array}{cc}
           s_+ & s_- \\
           s_- & s_+ \\
         \end{array}\right).
\end{equation}
Here $s_{\pm}=\left(1/2\right)\left(1 \pm \exp(i\Delta
\varphi_n)\right)$ and $\Delta
\varphi_n=\varphi_{n,1}-\varphi_{n,0}$, and $\varphi_{n,m}$ is
defined in Eq. (\ref{phi}). Neglecting high-oscillatory terms we
can write the Hamiltonian (\ref{ham1t}) in the rotating wave
approximation \cite{Louisell} as
\begin{equation}\label{ham1tt}
H_1=i\hbar g \left( e^{-i\omega t}\sigma ^ + s^{01} - h.c.\right).
\end{equation}

\subsection{Equations for the Wigner Functions}
 The Liouville-von Neumann equation for the density operator of the
electron $\rho_{mm',ss'}(x,x',t)$ is given by
\begin{equation}\label{Liouville}
    i\hbar \dot \rho=[H,\rho],
\end{equation}
where $s,s'=\pm$ are variables associated with spin degree of
freedom.  Henceforth  we omit the indices $n,n'$ due to confinement
in $z$ direction since we are interested only in the transitions
between the ground $m=0$ and $m=1$ transverse $y$ subbands. The
Wigner function can be obtained by integrating~\cite{Wigner}
\begin{equation}\label{Wignerf}
 W_{mm',ss'}(R,k,t)=\int \rho_{mm',ss'}(R,\Delta r,t) \exp(-i k \Delta
 r) d\Delta r,
\end{equation}
where the density operator is written in the new spatial variables
$R=(r+r')/2$ and $\Delta r=r-r'$ and $k$ is the electron wave
vector.

Following the standard procedure (see, e.g.,~\cite{Saikin}) we
derive a set of transport equations by neglecting non-local
correlations in the diagonal components of the Wigner functions
\begin{widetext}
\begin{eqnarray}\label{Wignersystemdiag}
\dot W_{11,++}  &+& v_{1,+} \frac{{\partial W_{11,++} }}{{\partial
x}} + \frac{{eE_x}}{\hbar }\frac{{\partial W_{11, ++ } }}{{\partial
k}} = g\left[ \left( s_+  W_{21, ++ }  + c.c. \right) + \left( s_-
W_{21, -+ }  + c.c. \right) \right], \\ \dot W_{11, - - } &+& v_{1,
- } \frac{{\partial W_{11, -  - } }}{{\partial x}} +
\frac{{eE_x}}{\hbar }\frac{{\partial W_{11, -  - } }}{{\partial k}}
= g\left[ \left( s_ -  W_{21, +  - }  + c.c. \right) + \left(
s_ + W_{21, -  - }  + c.c. \right) \right],\\
 \dot W_{22,--}&+&v_{2,-} \frac{\partial W_{22,--} }{\partial x} + \frac{eE_x}{\hbar
}\frac{\partial W_{22,--} }{\partial k} = -g\left[ \left( s_ -
W_{21, -+}  + c.c. \right) + \left( s_ + W_{21,-- }  + c.c.
\right) \right],\\ \dot W_{22,++}&+& v_{2,+} \frac{\partial
W_{22,++} }{\partial x} + \frac{eE_x}{\hbar }\frac{\partial
W_{22,++}}{\partial k} = -g\left[ \left( s_ +  W_{21, ++}  + c.c.
\right) + \left( s_ - W_{21,+- }  + c.c. \right) \right].
\end{eqnarray}
\end{widetext}
For the amplitudes of the off-diagonal components we define
$W'_{mm',ss'}(t)=W_{mm',ss'}(t)\exp(-i\omega t)$, and we obtain
\begin{widetext}
\begin{eqnarray}
\dot W'_{12, ++} + \frac{v_{1,+}+v_{2,+}}{2} \frac{\partial
W'_{12, +  + } }{\partial x} + \frac{{eE_x}}{\hbar
}\frac{{\partial W'_{12, + + } }}{{\partial k}} &-& i(\omega _{12,
+  + }  - \omega )W'_{12, + + }\\
 & =& g\left[ {s_ +  \left( {W_{22, +  + }  - W_{11, +  + } } \right) + s_ -  \left( {W_{22, -  + }  - W_{11, +  - } } \right)} \right],
 \nonumber\\
\dot W'_{12, +  - } + \frac{v_{1,+}+v_{2,-}}{2} \frac{{\partial
W'_{12, +  - } }}{{\partial x}} + \frac{{eE_x}}{\hbar
}\frac{{\partial W'_{12, + - } }}{{\partial k}} &-& i(\omega _{12,
+  - }  - \omega )W'_{12, + -}\\ &=& g\left[ {s_ +  \left(
{W'_{22, +  - }  - W'_{11, +  - } } \right) + s_ -  \left( {W_{22,
-  - }  - W_{11, +  + } } \right)} \right], \nonumber\\ \dot
W'_{12, -  - } + \frac{v_{1,-}+v_{2,-}}{2} \frac{{\partial W'_{12,
-  - } }}{{\partial x}} + \frac{{eE_x}}{\hbar }\frac{{\partial
W_{12, -  - } }}{{\partial k}} &-& i(\omega _{12, -  - }  - \omega
)W_{12, - -}\\ &=& g\left[ {s_ +  \left( {W_{22, -  - }  - W_{11,
-  - } } \right) + s_ - \left( {W'_{22, +  - } - W'_{11, -  + } }
\right)} \right], \nonumber\\
 \dot W'_{12, -  + } + \frac{v_{1,-}+v_{2,+}}{2} \frac{{\partial W'_{12, -  + } }}{{\partial x}} + \frac{{eE_x}}{\hbar }\frac{{\partial W'_{12, -  + } }}{{\partial k}} &-& i(\omega _{12, -  + }  - \omega )W'_{12, -  +}\\
 &=& g\left[ {s_ +  \left( {W'_{22, -  + }  - W'_{11, -  + } } \right) + s_ -  \left( {W_{22, +  + }  - W_{11, -  - } } \right)} \right], \nonumber\\
 \dot W'_{11, -  + }  + \frac{v_{1,-}+v_{1,+}}{2} \frac{{\partial W'_{11, -  + } }}{{\partial x}} + \frac{{eE_x}}{\hbar }\frac{{\partial W'_{11, -  + } }}{{\partial k}} &-& i(\omega _{11, -  + }  - \omega )W'_{11, -  + }\\
 &=& g\left[ {\left( {s_ -  W'_{21, +  + }  + s_ +  W'_{21, -  + } } \right) + \left( {s_ + ^* W'_{12, -  + }  + s_ - ^* W'_{12, -  - } } \right)} \right], \nonumber\\
 \dot W'_{22, -  + }  + \frac{v_{2,-}+v_{2,+}}{2} \frac{{\partial W'_{22, -  + } }}{{\partial x}} + \frac{{eE_x}}{\hbar }\frac{{\partial W'_{22, -  + } }}{{\partial k}} &-& i(\omega _{22, -  + }  - \omega )W'_{22, -  + }
 \label{Wignersystemoffdiag}\\
  &=&  - g\left[ {\left( {s_ - ^* W'_{12, +  + }  + s_ + ^* W'_{12, -  + } } \right) + \left( {s_ +  W'_{21, -  + }  + s_ -  W'_{21, -  - } } \right)}
  \right].
  \nonumber
\end{eqnarray}\newpage
\end{widetext}

For completeness we also included a static electric field $E_x$
along the $x$ direction, for example, due to a bias voltage.
$\omega_{mm',ss'}=\hbar^{-1}(E_{m',s'}-E_{m,s})$. The left part of
the equations (\ref{Wignersystemdiag}-\ref{Wignersystemoffdiag})
describes the ballistic transport of the electron in the quantum
wire, the right part is responsible for excitations induced by the
radiation. We consider the "ideal" contact boundary conditions for
a wire of length $L$
\begin{equation}\label{boundary1}
W_{mm,ss} (0,k)|_{v_{m,s}(k)>0}=f(k,\mu_l),
\end{equation}
\begin{equation}\label{boundary2}
W_{mm,ss} (L,k)|_{v_{m,s}(k)<0}=f(k,\mu_r).
\end{equation}
where $f(k,\mu)=1/(1+\exp[(E_{m,s}(k)-\mu)/(k_B T)])$ is the Fermi
function, $k_B$ is the Boltzmann constant, $T$ is the electron
temperature and $\mu_{l/r}$ are chemical potentials of the
left/right lead, respectively. We also assume that only internal
part of the quantum wire is irradiated, that is $E_y=0$ for $x<0$
and $x>L$.

 The electron charge density, electric
charge and spin currents can be obtained from Wigner functions as
\begin{equation}\label{density}
 n(x)= \frac{e}{2\pi} \sum_{m,s}\int_{-\infty}^{\infty} W_{mm,ss}(x,k)
 dk,
\end{equation}
\begin{equation}\label{current}
 I(x)= \frac{e}{2\pi} \sum_{m,s}\int_{-\infty}^{\infty}v_{m,s}(k) W_{mm,ss}(x,k)
 dk,
\end{equation}
and
\begin{eqnarray}\label{spincurrent}
 I_{\gamma}^{s}(x)= \frac{1}{2\pi} \sum_{m,s}\int_{-\infty}^{\infty}
 \langle \Psi_{m,s}| \sigma_\gamma | \Psi_{m,s}\rangle v_{m,s}(k) \nonumber \\W_{mm,ss}(x,k)
 dk,
\end{eqnarray}
respectively. Here $\gamma=(x,y,z)$ and different matrix elements
can be found in accordance with (\ref{phi}) as $\langle
\Psi_{m,\pm}| \sigma_x | \Psi_{m,\pm}\rangle=\pm \cos(\varphi_m)$,
$\langle \Psi_{m,\pm}| \sigma_y | \Psi_{m,\pm}\rangle=\mp \sin
(\varphi_m)$. The details of the numerical model and solution for
the derived system of equations are given in the next sections.

\section{Discrete Model}\label{discrete}
The form of Eqs.(\ref{Wignersystemdiag}-\ref{Wignersystemoffdiag})
does not allow us to solve the problem analytically even for the
stationary case $\partial  W/\partial t=0$ and unbiased channel
$E_x=0$. The solution is complicated by different inter-subband
transitions with the change of spin state $s$ originating from the
subband asymmetry in $k$ domain. This effect plays a central role
in the electric current generation and should be taken into
account. Thus, the system of equations
(\ref{Wignersystemdiag}-\ref{Wignersystemoffdiag}) was solved
numerically for $\bar y=25~\rm{nm}$, $\mu=0.0001~\rm{eV}$ and
$T=0.1K$. We model the domain $x\in [0,L]$ and $k\in[-k_{max},
k_{max}]$ with the mesh sizes of $\Delta x=L/(N_x-1)$ and $\Delta
k=2 k_{max}/(N_k-1)$, respectively. In the calculation we used
$N_x=25$ and $N_k=80$ and the length of the quantum wire
$L=2.5\;\mu\rm{m}$. The value $k_{max}=3.37\cdot10^7 \rm{m^{-1}}$
was chosen to ensure that all filled states in $k$ space are taken
into account. We fixed the values of diagonal components
$W_{mm,ss}$ on the boundaries at $x=0$ for $k$ with $v_{m,s}(k)>0$
and at $x=L$ for $k$ with $v_{m,s}(k)<0$, accordingly to
(\ref{boundary1},\ref{boundary2}). Similarly, for off-diagonal
components  we fixed the values $W_{m\neq m',s\neq s'}=0$ at $x=0$
for $k$ with $(v_{m,s}(k)+v_{m',s'}(k))>0$ and at $x=L$ for  $k$
with $(v_{m,s}(k)+v_{m',s'}(k))<0$."

The first order upwind difference scheme was used for the
propagation of the Wigner functions in $\{x,k\}$ domain and second
order two-step Lax-Wendroff scheme~\cite{Anderson} to describe the
time-dependent inter-subband transfer due to due to interaction
with the electro-magnetic field. The discretized Liouville equation
for the Wigner function can be written as
\begin{widetext}
\begin{eqnarray}\label{discretLiouvill}
&& W(x_i,k_j,t_{l+1/2})=W(x_i,k_j,t_{l})-\frac{\Delta
t}{2}\left[v(k_j)\frac{\Delta W(x_i,k_j,t_l)}{\Delta
    x}-F(W(x_i,k_j,t_l))\right],\\
 &&W(x_i,k_j,t_{l+1})=W(x_i,k_j,t_{l})-\Delta t\left[v(k_j)\frac{\Delta W(x_i,k_j,t_l)}{\Delta
    x}-F(W(x_i,k_j,t_{l+1/2}))\right]\nonumber
.\nonumber
\end{eqnarray}
\end{widetext}
where we consider the case $E_x=0$ and do not show subband indices
for brevity. The upwind/downwind difference is chosen in accordance
with
\begin{equation}\label{difference}
    \Delta W(x_i,k_j,t_l)=\left\{
\begin{array}{c}
     W(x_i,k_j,t_l)- W(x_{i-1},k_j,t_l) \\
\\
     {\rm if} \; v(k_j)>0, \\
\\
      W(x_{i+1},k_j,t_l)- W(x_i,k_j,t_l)\\
      \\
       {\rm if} \; v(k_j)<0.\\
\end{array}
\right.
\end{equation}
The function $F(W(x_i,k_j,t_l))$ embodies the remaining part of
Eqs.~(\ref{Wignersystemdiag}-\ref{Wignersystemoffdiag}) which
depends only on the Wigner function  $W(x_i,k_j,t_l)$ itself and
does not contain partial derivatives. The upwind differencing is
stable~\cite{Anderson} provided the time step is small enough:
$\Delta x/\Delta t\leq v_{max}$, where $v_{max}$ is the maximum
possible absolute value of the velocity. Additionally, the time
step must be much smaller than the highest frequency of the
solution. This condition is satisfied by $\Omega_R^{max} \Delta
t\ll1$ where $\Omega_R^{max}=\max\left\{\sqrt
{(\omega_{mm',ss'}-\omega)^2+4g^2}\right\}$ is the maximum
possible Rabi frequency involved in the problem. The  calculation
proved to be stable if these two conditions are met.

\begin{figure}[t]
\centering
\includegraphics[angle=270,width=8.5cm]{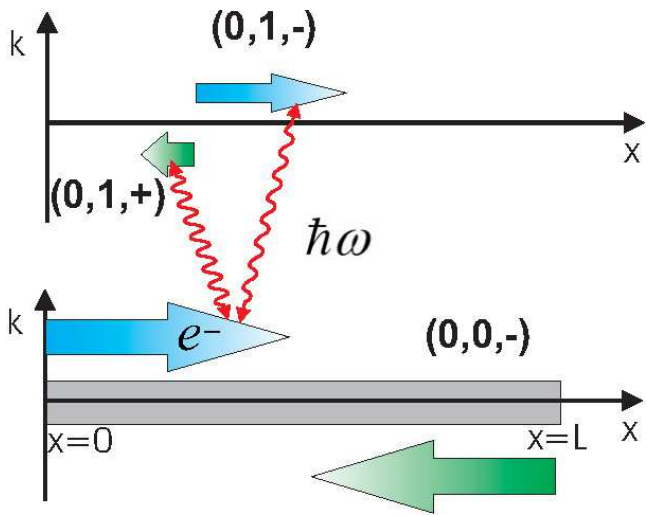}
\caption{(Color online) The different subband transition due to
the interaction with electro-magnetic wave are sketched. The
interaction causes some electrons to change the direction of the
velocity.} \label{fly}
\end{figure}

The investigation of the effects of the external bias and charge
redistribution in the quantum wire is not in the scope of this
paper. However, the way to include these effects is
straightforward. To consider the effect of voltage applied to the
quantum wire it suffices to add the $x$ component of the electric
field $E_x(x,V(t))$ in the discrete model (\ref{discretLiouvill}).
This electric field is a function of applied potential difference
$V(t)=(\mu_r-\mu_l)/e $ which can be time-dependent and
$x$-dependent. The latter is defined by the leads geometry.
Additional components  of electric field $E_x^q(n(x))$ can be
calculated self -consistently at each time step to incorporate the
effect of charge redistribution.

As was mentioned earlier, transitions between subbands can force
the electron to change the direction of the velocity. As a result
the flows of electrons moving in the opposite directions inside
the quantum wire intermix as shown in Fig.~\ref{fly}. Without
reflections in the wire, the steady-state solution can be obtained
by simply imposing the condition $\partial W/\partial t=0$ and
advancing from the given values at left/right boundaries for
$v(k)\gtrless k$. However, in the presence of radiation, the
electron distribution at the boundary acquires an additional
component, due to the electrons whose velocity direction is
changed, and the described above scheme fails. In order to achieve
the steady-state solution we considered the temporal dynamics of
the system evolving from some initial state until it reached
stationary conditions: $\partial W/\partial t\backsimeq0$ and
$I(x)\backsimeq I(x')$ for all $x,x'\in [0,L]$. As a initial state
we took the values of Wigner function at equilibrium $W_{mm',ss'}
(x,k)=\delta_{mm'}\delta_{ss'}f(k,\mu)$. This corresponds to the
uniformly distributed electron density along the channel and is a
solution of
Eqs.~(\ref{Wignersystemdiag}-\ref{Wignersystemoffdiag}) in the
absence of radiation and external bias: $E_x=E_y=0$. The chosen
method of obtaining the steady-state solution provides us also
with the transient behavior and, thus, gives more insight into the
problem. The drawback is a serious computational effort. The
electron distribution reaches the stationary state within the
effective time of flight through the quantum wire. Electrons
constantly change the direction of velocity due to the interaction
with electro-magnetic wave, therefore this time can be very long
compared to the time step $\Delta t$. Fortunately, the slower
electrons give the smaller contribution to the current and a
steady-state solution can be always found within certain accuracy.
The number of steps in time can reach values as much as
$N_t\sim10^5$. The results of the numerical simulation are
presented in the following section~\ref{sec4}.

\section{Results and Discussion} \label{sec4}

\begin{figure}[t]
\centering
\includegraphics[angle=270,width=8.5cm]{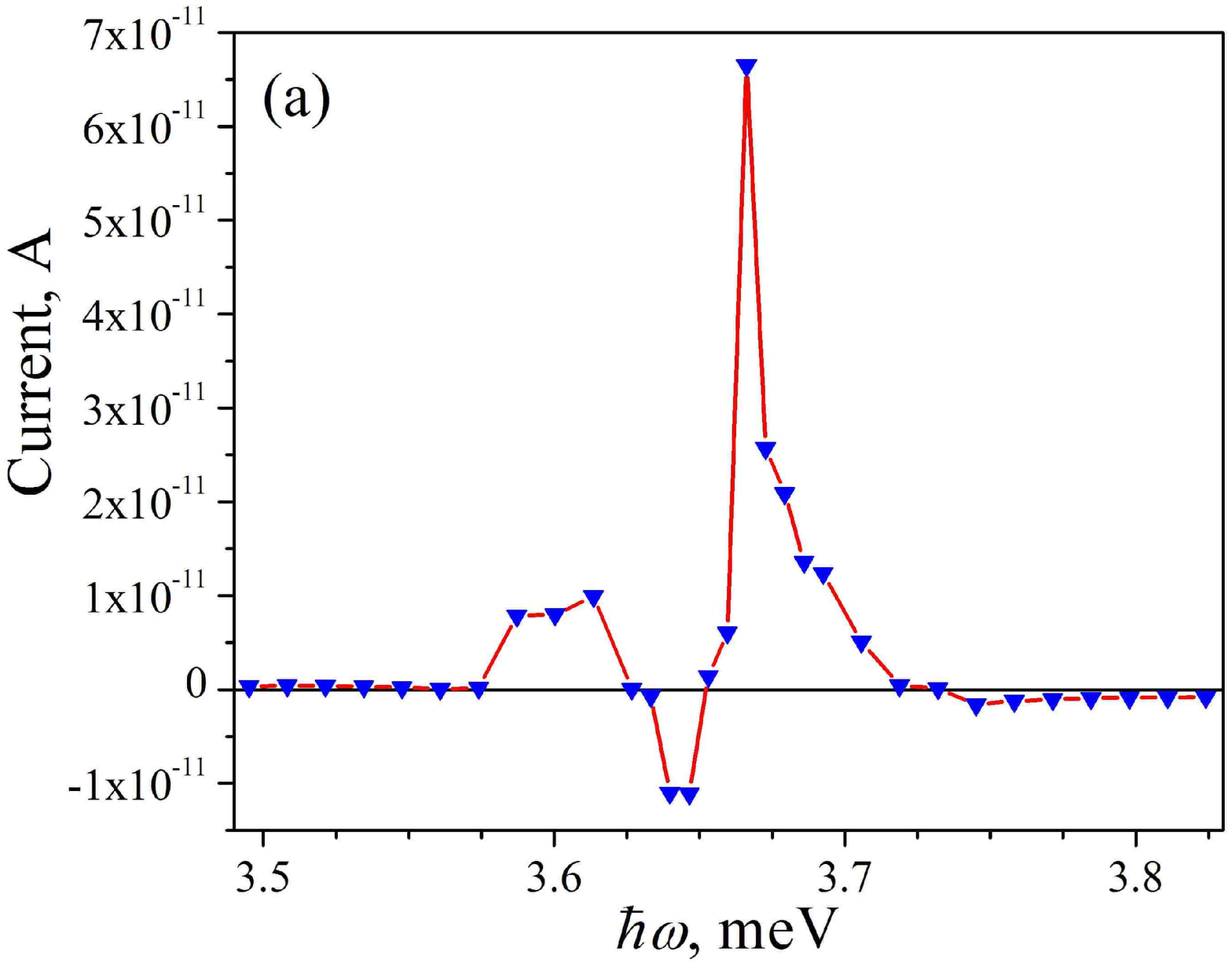}
\includegraphics[angle=270,width=8.5cm]{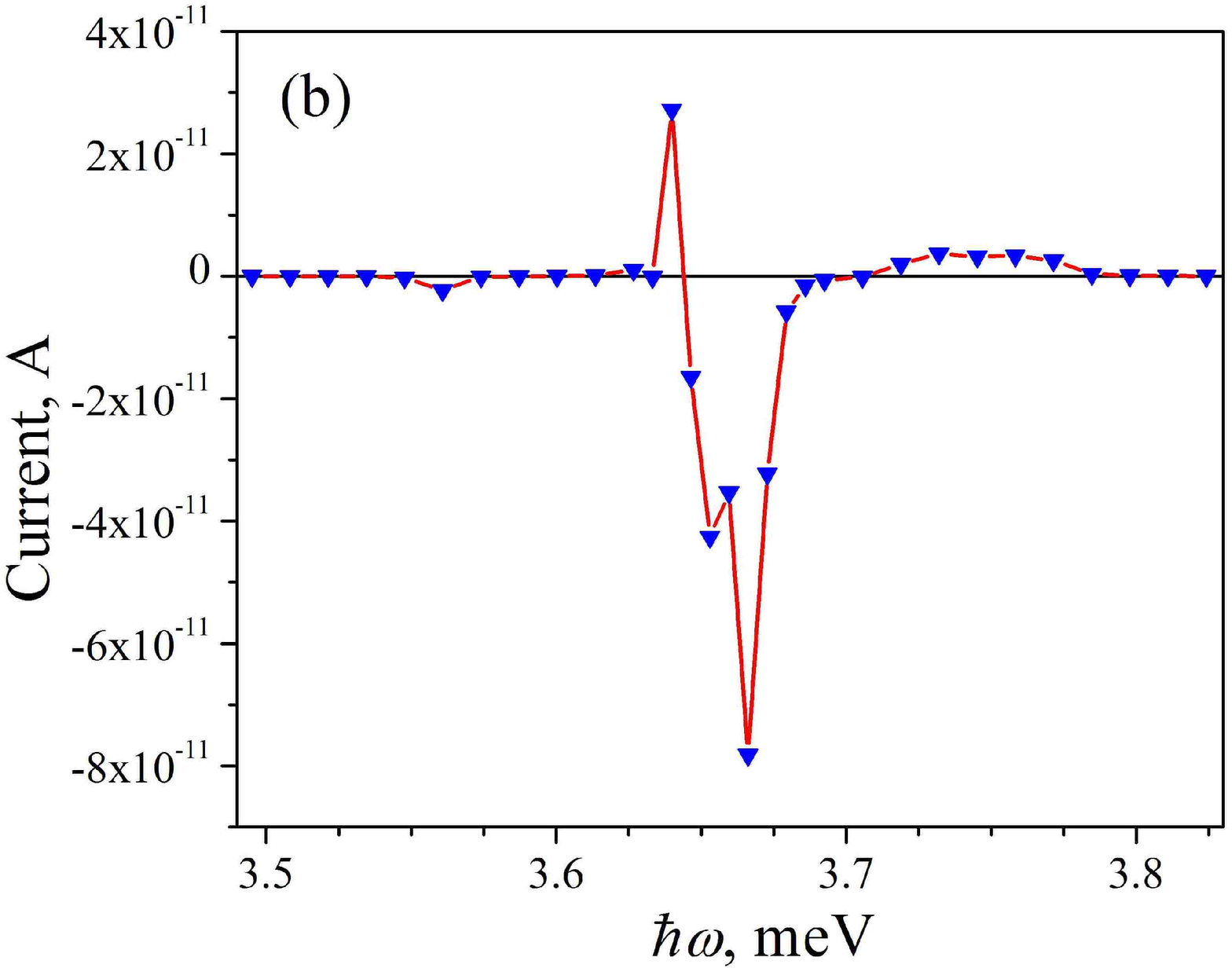}
\caption{(Color online) Current through the wire as a function of
photon energy $\hbar \omega$ for (a) $B_x=2$T, $B_y=2$T and (b)
$B_x=2$T, $B_y=-2$T. The parameters values are as in Fig.
\ref{fig2}, $E_y=200$V/m.} \label{fig3}
\end{figure}

The photo-induced current through the wire is shown in Fig.
\ref{fig3} as a function of the photon energy. The directions and
strength of the magnetic field in this graph are the same as in
Fig. \ref{fig2}. The amplitude of the electric field $E_y=200$V/m
used in our calculations was selected close to the electric field
amplitude used in recent experiments \cite{mani}. Fig. \ref{fig3}
clearly shows a number of current peaks corresponding to different
transitions. These peaks depend on the magnetic field strength and
direction, as a consequence of the magnetic field dependence of the
energy spectrum.

\begin{figure}[t]
\centering
\includegraphics[angle=0,width=8.5cm]{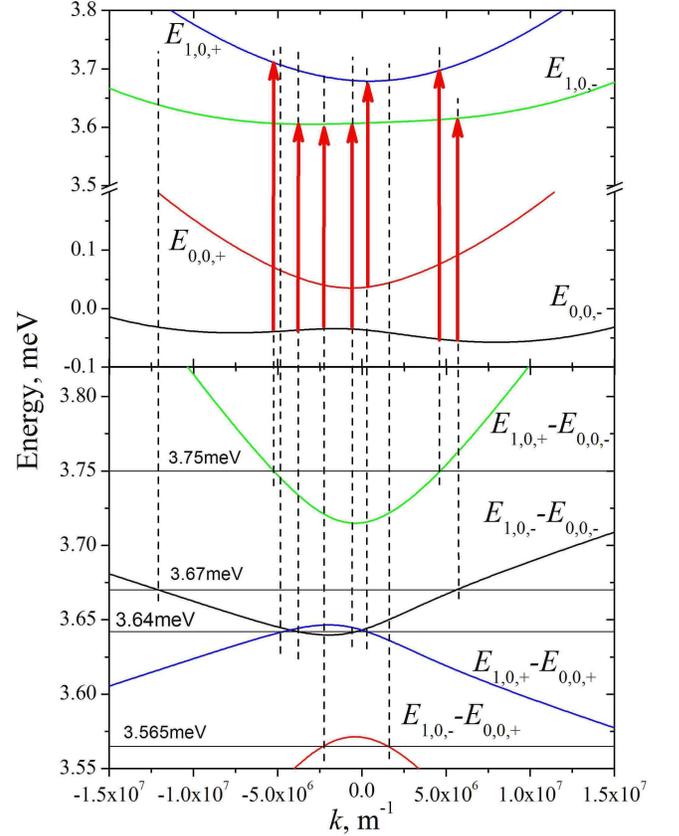}
\caption{(Color online) Transitions in the regions of photocurrent
peaks for $B_x=2$T, $B_y=-2$T. The energy spectrum and difference
energies are shown in the top and bottom panels, respectively.
Horizontal lines in the bottom panel correspond to excitation
energies in peak regions, vertical arrows in the top panel denote
transitions when electron velocity direction changes.}
\label{fig4}
\end{figure}

In order to understand transitions leading to a specific peak
formation, we consider in detail the current dependence on photon
energy for $B_x=2$T, $B_y=-2$T (Fig. \ref{fig3}(b)). It follows
from Fig. \ref{fig3}(b) that the current as a function of photon
energy has a well pronounced positive peak at $\hbar \omega \simeq
3.64$meV, a double negative peak with a minimum at $\hbar
\omega\simeq 3.67$meV, a small negative peak at $\hbar
\omega\simeq 3.565$meV, and a broad positive peak of small
amplitude at $\hbar \omega\simeq 3.75$meV. Fig. \ref{fig4}
represents a graphical determination of relevant transitions.

The energy spectrum of two lowest spin-splitted subbands is
plotted in the top panel of Fig. \ref{fig4}. The bottom panel of
Fig. \ref{fig4} shows the energy difference between different
transverse subbands. By plotting horizontal lines corresponding to
the peak energies, in the bottom panel and, by drawing vertical
lines through the intersection points of those horizontal lines
with energy difference, we can finallyidentify the points in
the top panel corresponding to the peak formation. As it was
mentioned above, the important transitions are those that lead to
a change of the electron velocity direction. These transitions are
shown by arrows in the top panel of Fig. \ref{fig4}.

In particular, let us consider the large negative peak in the
current at $\hbar \omega\simeq 3.67$meV (Fig. \ref{fig3}(b)). Fig.
\ref{fig4} shows that the horizontal line $3.67$meV intersects
only $E_{1,0,-}-E_{0,0,-}$ curve in two points. We remind that the
electron velocity is determined by the slope of $E_{m,n,\pm}(k)$
according to Eq. (\ref{velocity}). The left intersection point
gives $k$-vector of transition in which the electron velocity
direction is conserved), because the slopes of $E_{1,0,-}$ and
$E_{0,0,-}$ at this value of $k$ are in the same direction. The
right intersection point of $3.67$meV line with
$E_{1,0,-}-E_{0,0,-}$ curve gives a transition with a change of
electron velocity direction, specifically, with a back-scattering
of left-moving electrons. Consequently, the electron flux from the
right to the left decreases and, because of the negative electron
charge, a negative current appears.

\begin{table}[b]
\begin{tabular}{c|c}
Excitation energy, meV & Transitions \\ \hline 3.565 &
$E_{0,0,+}\rightarrow E_{1,0,-}$\\
 3.64 &
$E_{0,0,-}\rightarrow E_{1,0,-}$; $E_{0,0,+}\rightarrow
E_{1,0,+}$\\ 3.67 & $E_{0,0,-}\rightarrow E_{1,0,-}$\\
 3.75 & $E_{0,0,-} \rightarrow
E_{1,0,+}$
\end{tabular}
\caption{Transitions giving contribution to photocurrent at
selected radiation frequencies for $B_x=2$T, $B_y=-2$T.}
\label{table1}
\end{table}

Similarly, one can consider transitions at other radiation
frequencies. An interesting situation occurs for $\hbar \omega \simeq
3.64$meV excitation, since at this particular frequency three out of
four transitions are characterized by the reverse of electron velocity
direction. We summarize transitions contributing to the photocurrent
at selected radiation frequencies in Table \ref{table1}. The same
analysis can also be applied to the result presented in
Fig. \ref{fig3}(a), but, because of a more distorted spectrum, the
roles of different possible transitions are more difficult to
interpret.. Moreover, we would like to note that, generally,
transition probabilities from $E_{0,0,-}$ to $E_{1,0,+}$ and from
$E_{0,0,+}$ to $E_{1,0,-}$ subbands are smaller than transition
probabilities from $E_{0,0,-}$ to $E_{1,0,-}$ and from $E_{0,0,+}$ to
$E_{1,0,+}$ subbands because of the different spin direction in
initial and final states. This results in a smaller current peaks at
$\hbar \omega\simeq 3.565$meV and $\hbar \omega\simeq 3.75$meV in
Fig. \ref{fig3}(b).

\begin{figure}[t]
\centering
\includegraphics[angle=270,width=8.5cm]{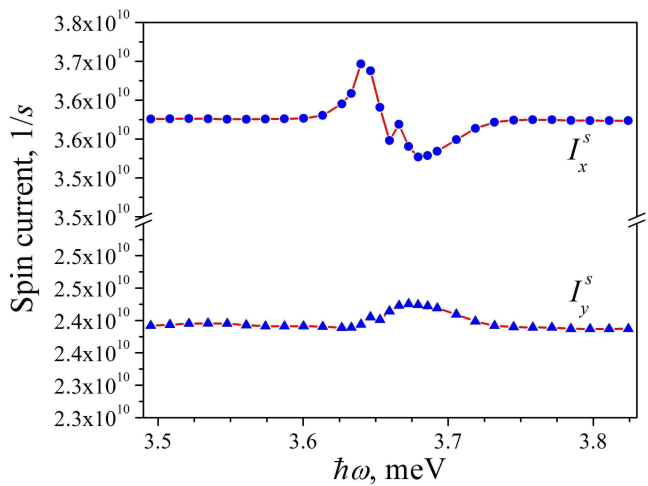}
\caption{(Color online) Spin current components for $B_x=2$T,
$B_y=-2$T at $x=0.$} \label{fig5}
\end{figure}

As electrons carry spin as well as charge, the external radiation also
changes the spin current through the wire. Notice that even without
the radiation, the spin current is not zero, due to the spin-orbit
interaction. Fig. \ref{fig5} shows spin current components at $x=0$
for $B_x=2$T, $B_y=-2$T. We note that the spin current dependence on
the radiation frequency has features similar to the charge current
(Fig. \ref{fig3}(b)). However, we found that radiation-induced changes
in spin current are much less than the equilibrium spin current in the
wire. From an experimental point of view, the spin currents are not so
easy detectable. Therefore, the observation of this photovoltaic
effect through spin current seems unpractical.

\section{Conclusions} \label{sec5}

In summary, in this paper we have investigated the photovoltaic effect
in quantum wires with spin-orbit interaction and in-plane magnetic
field. We have found that the peculiarities of the energy spectrum
lead to a photocurrent generation. The dependence of the photoinduced
current on the excitation frequency was calculated numerically using
the Wigner functions formalism. A system of coupled equations for the
Wigner functions was derived and solved numerically for "ideal"
contact boundary conditions. We used the first order upwind
differencing for the propagation in the spatial domain and the second
order two-step Lax-Wendroff differencing for time-dependent
inter-subband transitions due to electro-magnetic wave
excitation. Stable numerical solutions were found under appropriate
choices of the time-step $\Delta t$. The calculations can be extended
to introduce the effects of an external bias and self-consistent
potentials due to charge density redistributions which can be a topic
for a future investigation. The frequency dependence of the
photoinduced current consists of a set of peaks related to transitions
between different points of the spectrum.  Therefore, the energy
spectrum can be reconstructed from photocurrent measurements. Material
parameters, such as spin-orbit coupling constants, can be obtained
from the analysis of the photocurrent.

\section*{Acknowledgments}

We gratefully acknowledge useful discussions with M. Cheng and L.
Fedichkin. This research was supported by the National Science
Foundation, Grants DMR-0121146 and DMR-0312491.

\end{document}